\documentclass[12pt]{article}
\usepackage{epsf}

\makeatletter
\@addtoreset{equation}{section}

\makeatother

\newcommand{\gbar}{\bar{g}}
\newcommand{\ghat}{\hat{g}}
\newcommand{\Rhat}{\hat{R}}
\newcommand{\NP}[1]{{\it Nucl.\ Phys.\ }{\bf #1}}
\newcommand{\PL}[1]{{\it Phys.\ Lett.\ }{\bf #1}}

\newcommand{\PR}[1]{{\it Phys.\ Rev.\ }{\bf #1}}
\newcommand{\PRL}[1]{{\it Phys.\ Rev.\ Lett.\ }{\bf #1}}

\begin{document}

\thispagestyle{empty}
%
\begin{flushright}
TIT/HEP--461 \\
{\tt hep-th/0101042} \\
January, 2001 \\
\end{flushright}
\vspace{3mm}
\begin{center}
{\Large
{\bf  Our World as an Intersection of Walls and a String
 }} 
\\[18mm]

{\sc Norisuke~Sakai}\footnote{
\tt e-mail: nsakai@th.phys.titech.ac.jp
} \hspace{2.0mm}
and \hspace{2.0mm}
{\sc Shigemitsu~Tomizawa}\footnote{
\tt e-mail: stomizaw@th.phys.titech.ac.jp
} \\[3mm]
{\it Department of Physics, 
Tokyo Institute of Technology \\[2mm]
Oh-okayama, Meguro, Tokyo 152-8551, Japan} \\[4mm]

%
\vspace{18mm}
{\bf Abstract}\\[5mm]
{\parbox{13cm}{\hspace{5mm}
A solution of Einstein equations is obtained for our four-dimensional world 
as an intersection of a wall and a string-like defect in 
seven-dimensional spacetime with a negative cosmological constant. 
A matter energy-momentum tensor localized 
on the wall and on the string is needed. 
A single massless graviton is found and is localized around the 
intersection. 
The leading correction to the gravitational Newton potential 
from massive spin 2 graviton is found to be almost identical 
to that of a wall in five dimensions, contrary to 
the case of a string in six dimensions. 
The  generalization to the intersection of a string and $n$ 
orthogonally intersecting walls is also obtained 
and a similar result is found for the gravitational 
 potential. 
}}
\end{center}
\vfill
\newpage
\setcounter{page}{1}

\section{Introduction}\label{INTRO}

{}In recent years, a lot of attention has been paid to topological objects 
 such as domain walls \cite{AbrahamTownsend}, \cite{DW}, junctions 
\cite{DWJ}--\cite{junction} in field theories, and  branes in string theories 
\cite{D-brane}. 
An interesting idea has been advocated that 
our four-dimensional world may be realized on these topological objects 
embedded in higher dimensional spacetime \cite{ADD}. 
In this ``brane world" scenario, 
most of the particles in the standard model should be obtained as 
modes localized on the topological defects. 
These models are based on the fact that the Newton's law for the 
gravitational force is tested only at short distances of the order
of a hundred microns \cite{Price}. 
In view of the possibility of a fundamental theory in higher 
dimensions such as the ten-dimensional superstring, it is desirable 
to have a model with more extra dimensions and with supersymmetry. 
In fact, a model for supersymmetry breaking has been proposed 
based on the coexistence of branes \cite{MSSS}. 

Another fascinating possibility has also been proposed 
to consider walls in five-dimensional spacetime with a 
negative cosmological constant \cite{ph/9905221}. 
It has been recognized that the graviton in our four-dimensional world 
can be obtained as a zero mode localized on the wall embedded in the 
five-dimensional curved spacetime \cite{th/9906064}. 
The model has been extended to intersections of walls in 
higher dimensional AdS spacetime \cite{th/9907209}. 
Topological defects of codimension two \cite{Chodos-Poppitz}, 
\cite{th/0004014} and more \cite{th/0006251} 
 have also been studied. 
{}Following the Kaluza-Klein idea, the graviton field in the bulk 
is decomposed into spin 2 graviton, spin 1 graviphoton, 
and spin 0 graviscalar fields in four-dimensional spacetime. 
It has been shown that the spin 0 field becomes unphysical for an isolated 
wall, but becomes a physical scalar field 
called radion if there are two or more walls \cite{th/9912160}. 
The graviphoton has also been studied in the case of walls in 
five dimensions \cite{th/0010208}. 
However, these scalar and vector fields are not well studied in the 
case of topological defects in higher dimensional spacetime. 
The contributions from the spin 2 gravitons to the Newton potential 
for the gravitational force have been obtained, but the contributions 
from spin 1 and spin 0 fields are still an open question in these 
higher dimensional models. 
A number of works also explored the possible extensions of the model 
to higher extra dimensions \cite{th/0006203}. 
It is important to study the coexistence of these topological defects 
in higher dimensional spacetime to build a model from a fundamental 
theory with higher spacetime dimensions and/or higher supersymmetry 
 \cite{MSSS}. 

The purpose of this paper is to construct a model for our four-dimensional 
world as an intersection of a wall (codimension one) and a string-like 
defect (codimension two)
and to obtain corrections to the gravitational Newton 
potential by working out the spin 2 modes on the background. 
We find a configuration of a four-dimensional intersection 
 of a wall and a string-like defect in 
seven-dimensional spacetime. 
We obtain a four-dimensional graviton as a zero mode localized 
on the intersection. 
It turns out necessary to have a matter energy-momentum tensor besides the 
cosmological constant on the topological defects. 
The correction to the Newton's law is found to be similar to that of 
the wall in five dimensions contrary to the string-like defect 
in six dimensions. 
We also work out generalizations to intersections of 
a string-like defect and $n$ orthogonally intersecting walls 
in ($6+n$) dimensional spacetime. 
A single massless graviton is found to be localized at the intersection 
and the leading correction to the Newton potential 
is found to be similar to the single wall case. 
We leave the study of the spin 1 and spin 0 fields for a future work. 

In sect.~2, we will study the solution of the Einstein equation 
representing the intersection of walls and a string-like defect 
embedded in seven-dimensional spacetime. 
The modes on the background of intersection of walls and 
a string-like defect are obtained, and the correction for the 
gravitational Newton potential is worked out in sect.~3. 
In sect.~4 the extension of the solution is obtained 
for the intersection of a string-like defect and $n$ orthogonally 
intersecting walls in higher dimensional spacetime. 

\section{Intersection of a Wall and a String-like Defect 
} 
\label{WS:INTERSECTION1}
In order to consider an intersection of parallel walls and a string-like 
defect, we will work on a seven-dimensional spacetime. 
The indices for the bulk spacetime coordinates are 
denoted by capital alphabets $A, B, \cdots$ throughout this paper. 
{}Following Ref.\cite{ph/9905221}, we will compactify one of the extra 
dimensions and make an orbifold $S^1/Z_2$ by identifying points under the 
reflection : $y \rightarrow -y$. 
Two parallel five-branes (walls) with vanishing widths 
are placed at the fixed points of the orbifold: $y=0$ and $y=y_c$. 
The polar coordinates $(r, \theta)$ 
for the two more extra dimensions perpendicular to 
the $y$ direction take values $0 \le r < \infty$, and $0 \le \theta < 2\pi$. 
Perpendicular to the walls lies a four-brane (string-like defect) 
 at $r=0$. 
We will consider an infinitesimal thickness of radial 
size $\varepsilon$ for the string like defect following Ref.\cite{th/0004014}. 
Denoting the metric and scalar curvature in seven dimensions by $\hat{g}$ 
and $\hat{R}$ respectively, 
the action is given by 
\begin{eqnarray}
S &=& \int d^7x \, \sqrt{\hat{g}} 
\left( - \frac{1}{2} M_7{}^5 \hat{R} - \Lambda_7 \right) 
+ 
 \int_{y=0} d^6x \, \sqrt{-\bar{g}^{(1)}} 
\left( {\cal{L}}_{{\rm wall}}^{(1)} - \Lambda_6^{(1)} \right) 
\nonumber \\
& & {} +
\int_{y=y_c} d^6x \, \sqrt{-\bar{g}^{(2)}} 
\left( {\cal{L}}_{{\rm wall}}^{(2)} - \Lambda_6^{(2)} \right) 
+
\int d^7x \, \sqrt{\hat{g}} {\cal{L}}_{{\rm string}}  ~. 
\label{WS:action1}
\end{eqnarray}
where $M_7$ is the seven-dimensional Planck scale, $\Lambda_7$ is a bulk
cosmological constant and $\Lambda_6^{(1)}, \Lambda_6^{(2)}$ are tensions of 
five-branes (walls) which lie at $y=0, y_c$ respectively. 
The matter part of the Lagrangian for the walls are denoted as 
${\cal{L}}_{{\rm wall}}^{(1)}, {\cal{L}}_{{\rm wall}}^{(2)}$ and that 
for the string-like defect as ${\cal{L}}_{{\rm string}}$. 
Here we explicitly separated out the wall tensions 
 $\Lambda_6^{(1)}, \Lambda_6^{(2)}$ from the matter 
Lagrangians on the walls, but we included the string tension in the
Lagrangian ${\cal L}_{\rm string}$ of the string-like defect.
The induced metrics on the five-branes are denoted by $\bar{g}$ 
\begin{eqnarray}
 \gbar_{A B}^{(1)} \equiv \ghat_{A B}|_{y=0} ~, \quad 
\gbar_{A B}^{(2)} \equiv \ghat_{A B}|_{y=y_c}  ~,
\end{eqnarray}
for indices $A, B = 0,1,2,3,r,\theta$. 
Indices for the coordinates of our four-dimensional spacetime are denoted 
by Greek alphabets as 
 $x^\mu=0, \cdots, 3$. 

We will look for a solution that respects four-dimensional 
Poincar\'e invariance in the intersection. 
Let us assume the following ansatz for the seven-dimensional metric 
\begin{eqnarray}
ds^2 &=& \hat{g}_{AB} dx^A dx^B \nonumber\\
     &=& \sigma(y,r) \eta_{\mu \nu} dx^\mu dx^\nu - dy^2 - dr^2 - \gamma(y,r) d\theta^2  ~.
\label{WS:metric1}
\end{eqnarray}
This metric has two kinds of warp factors $\sigma(y,r)$ and $\gamma(y,r)$.
The seven-dimensional Einstein equations for the action~(\ref{WS:action1}) 
is given by 
\begin{eqnarray}
 \Rhat_{AB} &\!\!\!-&\!\!\! \frac{1}{2} \ghat_{AB} \Rhat
=
 \frac{1}{M_7{}^5} \Bigg[ \Lambda_7 \ghat_{AB} 
+ \hat{T}^{(\rm string)}_{AB}
+ \left(\bar{T}^{({\rm wall} 1)}_{AB} + \Lambda_6^{(1)} \right) 
\frac{\sqrt{-\bar{g}}^{(1)}}{\sqrt{\ghat}} \gbar_{A B}^{(1)}  \delta(y) 
\nonumber \\
&\!\!\! + &\!\!\! {} 
 \left(\bar{T}^{({\rm wall} 2)}_{AB} + \Lambda_6^{(2)} \right) 
\frac{\sqrt{-\bar{g}}^{(2)}}{\sqrt{\ghat}} \gbar_{A B}^{(2)} 
 \delta(y-y_c)  \Bigg] ~,
\quad A, B = 0, \cdots, 3, r, \theta, 
\label{WS:Einsteineq1}
\end{eqnarray}
if the indices $A,B$ do not contain $y$ component. 
If at least one of the indices $A, B$ corresponds to the $y$ component, 
the Einstein equation becomes 
\begin{eqnarray}
 \Rhat_{AB} - \frac{1}{2} \ghat_{AB} \Rhat
&\!\!\!=&\!\!\!
 \frac{1}{M_7{}^5} \Bigg[ \Lambda_7 \ghat_{AB} 
 + \hat{T}^{(\rm string)}_{AB}
  \Bigg] ~, 
\quad A \; {\rm or} \; B = y ~.
\label{WS:Einsteineq2}
\end{eqnarray}
The energy-momentum tensors for the string-like defect and the walls 
are defined as 
\begin{eqnarray}
 \hat{T}^{(\rm string)}_{AB} 
&\!\!\! 
\equiv 
&\!\!\! 
{2 \over \sqrt{\hat{g}}}
{\partial \left( \sqrt{\hat{g}} {\cal L}_{\rm string} \right)
\over \partial \hat{g}^{AB}}, 
\nonumber \\
\bar{T}^{({\rm wall} 1)}_{AB} 
&\!\!\! 
\equiv 
&\!\!\! 
{2 \over \sqrt{-\bar{g}^{(1)}}}
{\partial \left( \sqrt{-\bar{g}} {\cal L}_{\rm wall}^{(1)} \right)
\over \partial \bar{g}^{(1)}{}^{AB}} ~, 
\quad 
\bar{T}^{({\rm wall} 2)}_{AB} 
\equiv 
{2 \over \sqrt{-\bar{g}^{(2)}}}
{\partial \left( \sqrt{-\bar{g}^{(2)}} {\cal L}_{\rm wall}^{(2)} \right)
\over \partial \bar{g}^{(2)}{}^{AB}} ~.
\label{WS:enrgy-momentum}
\end{eqnarray}
Let us note that the ratios $\sqrt{-\gbar^{(1)}}/\sqrt{\ghat}$ and
$\sqrt{-\bar{g}^{(2)}}/\sqrt{\ghat}$ at the five-branes in 
Eq.(\ref{WS:Einsteineq1}) reduce 
to $1$ for the metric~({\ref{WS:metric1}}).

The non-zero components of the energy-momentum tensor from the matter
sources on the  wall at $y=0$ are assumed as 
\begin{eqnarray}
\bar{T}^{({\rm wall} 1) }{}^\mu_\nu = t_0^{(1)}(r) \delta^\mu_\nu ~, \quad
\bar{T}^{({\rm wall} 1) }{}^r_r = t_r^{(1)}(r)  ~, \quad
\bar{T}^{({\rm wall} 1) }{}^\theta_\theta = t_\theta^{(1)}(r) ~,
\label{WS:emt}
\end{eqnarray}
where the $r$-dependence of $t_0^{(1)}, t_r^{(1)}, t_\theta^{(1)}$ 
will be given later. 
A similar form is assumed for the energy-momentum tensor 
$\bar T^{({\rm wall} 2)}{}^A_B$ on the other  wall at $y=y_c$. 
On the other hand, we assume that the matter source for the string-like 
defect is distributed continuously within the core of radius $\varepsilon$ and 
vanishes outside of it.
The non-zero components of the energy-momentum tensor are assumed as 
\begin{eqnarray}
& &
\hat{T}^{(\rm string)}{}^\mu_\nu = f_0(y,r) \delta^\mu_\nu  ~, \quad
\hat{T}^{(\rm string)}{}^y_y = f_y(y,r)  ~, \quad
\hat{T}^{(\rm string)}{}^y_r = f_{yr}(y,r)  ~, \quad
\nonumber \\
& &
\hat{T}^{(\rm string)}{}^r_r = f_r(y,r)  ~, \quad
\hat{T}^{(\rm string)}{}^\theta_\theta = f_\theta(y,r)  ~,
\label{WS:emf}
\end{eqnarray}
where the $(y, r)$-dependence of $f_0, f_y, f_{yr}, f_r, f_\theta$ 
will be given later. 

Using the ansatz of the metric (\ref{WS:metric1}) 
and the energy-momentum tensors (\ref{WS:emt}) (\ref{WS:emf}), 
the Einstein equations~(\ref{WS:Einsteineq1}), (\ref{WS:Einsteineq2}) 
now become
\begin{eqnarray}
 &\!\!\! &\!\!\! \frac{3}{2}\frac{\sigma''}{\sigma} 
+ \frac{3}{4}\frac{\sigma'}{\sigma}\frac{\gamma'}{\gamma} 
+ \frac{1}{2}\frac{\gamma''}{\gamma} - \frac{1}{4}\frac{\gamma'^2}{\gamma^2} 
+ \frac{3}{2}\frac{\ddot{\sigma}}{\sigma} 
+ \frac{3}{4} \frac{\dot{\sigma}}{\sigma}\frac{\dot{\gamma}}{\gamma} 
+ \frac{1}{2}\frac{\ddot{\gamma}}{\gamma} 
- \frac{1}{4}\frac{\dot{\gamma}^2}{\gamma^2}  \nonumber\\ 
 &\!\!\!=&\!\!\! - \frac{1}{M_7{}^5} \left[ \Lambda_7 + f_0(y,r) 
+ (t_0^{(1)}(r) + \Lambda_6^{(1)})\delta(y) 
+ (t_0^{(2)}(r) + \Lambda_6^{(2)})\delta(y-y_c) \right] ~,
\label{WS:Einsteinmunu} \\ \nonumber \\ 
&\!\!\! &\!\!\! \frac{3}{2}\frac{\sigma'^2}{\sigma^2} 
+ \frac{\sigma'}{\sigma}\frac{\gamma'}{\gamma} 
+ 2\frac{\ddot{\sigma}}{\sigma} + \frac{1}{2}\frac{\dot{\sigma^2}}{\sigma^2} 
+ \frac{\dot{\sigma}}{\sigma}\frac{\dot{\gamma}}{\gamma} 
+ \frac{1}{2}\frac{\ddot{\gamma}}{\gamma} 
- \frac{1}{4}\frac{\dot{\gamma}^2}{\gamma^2}  
=
- \frac{1}{M_7{}^5} \left[ \Lambda_7 + f_y(y,r) \right]
 ~,
\label{WS:Einsteinyy} \\ \nonumber \\
&\!\!\! &\!\!\! \frac{\sigma'}{\sigma}\frac{\dot{\sigma}}{\sigma} 
- 2\frac{\dot{\sigma}'}{\sigma} 
+ \frac{1}{4}\frac{\gamma'}{\gamma}\frac{\dot{\gamma}}{\gamma} 
- \frac{1}{2}\frac{\dot{\gamma}'}{\gamma}  
 = - \frac{1}{M_7{}^5}f_{yr}(y,r)
 ~,
\label{WS:Einsteinyr} \\ \nonumber \\
&\!\!\! &\!\!\! 2\frac{\sigma''}{\sigma} 
+ \frac{1}{2}\frac{\sigma'^2}{\sigma^2} 
+ \frac{\sigma'}{\sigma}\frac{\gamma'}{\gamma} 
+ \frac{1}{2}\frac{\gamma''}{\gamma} 
- \frac{1}{4}\frac{\gamma'^2}{\gamma^2} 
+ \frac{\dot{\sigma}}{\sigma}\frac{\dot{\gamma}}{\gamma} 
+ \frac{3}{2}\frac{\dot{\sigma}^2}{\sigma^2} \nonumber \\
 &\!\!\!=&\!\!\! - \frac{1}{M_7{}^5} \left[ \Lambda_7 + f_r(y,r) 
+ (t_r^{(1)}(r) + \Lambda_6^{(1)})\delta(y) 
+ (t_r^{(2)}(r) + \Lambda_6^{(2)})\delta(y-y_c) \right]
 ~,
\label{WS:Einsteinrr} \\ \nonumber \\
&\!\!\! &\!\!\! 2\frac{\sigma''}{\sigma} 
+ \frac{1}{2}\frac{\sigma'^2}{\sigma^2} 
+ 2\frac{\ddot{\sigma}}{\sigma} + \frac{1}{2}\frac{\dot{\sigma}^2}{\sigma^2}
\nonumber \\
 &\!\!\!=&\!\!\! - \frac{1}{M_7{}^5} \left[ \Lambda_7 + f_\theta(y,r) 
+ (t_\theta^{(1)}(r) + \Lambda_6^{(1)})\delta(y) 
+ (t_\theta^{(1)}(r) + \Lambda_6^{(2)})\delta(y-y_c) \right]
 ~,
\label{WS:Einsteintt}
\end{eqnarray}
where the dash $(\, ' \,)$ and the dot $(\, \dot{} \,)$ denote partial
differentiations $\partial_y$ and $\partial_r$ respectively.

The boundary conditions at $r=0$ are assumed as \cite{th/0004014}
\begin{equation}
 \partial_r \sigma |_{r=0} = 0  ~, \quad 
 \partial_r \sqrt{\gamma} |_{r=0} = 1  ~, \quad
 \gamma |_{r=0} = 0  .
\label{WS:stringboundary}
\end{equation}
These conditions are consistent with the usual regular solution in flat
space so that $r=0$ can be regarded as a regular origin of the polar
coordinates ($r,\theta$). 

We assume that the solutions 
outside the core of the string-like defect 
($r>\varepsilon$) take the form 
\begin{eqnarray}
 \sigma(y, r) = e^{-a|y|-br}  ~, \quad \gamma(y, r) = R_0{}^2 e^{-c|y|-dr} ~,
\label{WS:metricansatzabcd}
\end{eqnarray}
which is consistent with the orbifold symmetry $y \leftrightarrow -y$ .
The coefficient $R_0$ is a constant for the length scale 
and $a$, $b$, $c$, $d$ are
constants with the dimension of mass. 
We assume that the warp factors do not diverge as $r$ goes to infinity  
\begin{equation}
b \ge 0, \qquad d \ge 0 ~.
\end{equation}

Substituting Eq.(\ref{WS:metricansatzabcd}) into the Einstein equations 
(\ref{WS:Einsteinmunu})--(\ref{WS:Einsteintt}),
we obtain solutions
\begin{eqnarray}
 c = \pm 2b ~,\quad d= \mp 2a ~,
\label{WS:solution1}
\end{eqnarray}
where the upper and lower signs are for $a \le 0$ and $a \ge 0$ respectively.
The constants $a$ and $b$ are related by 
\begin{equation}
 a^2 + b^2 = \frac{-2 \Lambda_7}{5 M_7{}^5}  ~,
\end{equation}
which means that this solution requires $\Lambda_7 < 0$.
Here we have one parameter family of solutions that are locally related by
an $SO(2)$ rotation of $(y,r)$-space, apart from the range of $(y,r)$ that is covered.

{}From the boundary conditions at $y = 0, y_c$, the components of the
energy-momentum tensor from the  walls for $r > \varepsilon$ must be
\begin{eqnarray}
 t_0^{(1)}(r) &\!\!\!=&\!\!\!  
 M_7{}^5 (3a \pm 2b)  - \Lambda_6^{(1)}  
~,
\nonumber \\
 t_r^{(1)}(r) &\!\!\!=&\!\!\!  
 M_7{}^5 (4a \pm 2b) - \Lambda_6^{(1)}  
~,
\nonumber \\
 t_\theta^{(1)}(r) &\!\!\!=&\!\!\! 
4  M_7{}^5 a - \Lambda_6^{(1)} ~,
\label{WS:emsolution1}
\end{eqnarray}
\begin{eqnarray}
 t_0^{(2)}(r) &\!\!\!=&\!\!\!  
   M_7{}^5 (-3a \mp 2b) - \Lambda_6^{(2)}    
~,
\nonumber \\
 t_r^{(2)}(r) &\!\!\!=&\!\!\!  
   M_7{}^5 (-4a \mp 2b) - \Lambda_6^{(2)}    
~,
\nonumber \\
 t_\theta^{(2)}(r) &\!\!\!=&\!\!\! 
  -4  M_7{}^5 a - \Lambda_6^{(2)}  ~.
\label{WS:emsolution2}
\end{eqnarray}
Therefore 
we can find a solution for $a$ and $b$ only 
if the energy-momentum tensors $t^{(1)}, t^{(2)}$'s from matter sources 
are present on the  walls. 
It is also interesting to observe that the matter energy-momentum tensors 
$t^{(1)}, t^{(2)}$'s on the walls are independent of the position 
$(r, \theta)$ on the wall outside the intersection region 
$r > \varepsilon$. 

Let us examine the Einstein equations near the core of the string-like 
defect $0 \le r< \varepsilon$. 
The components of the string tension are obtained by integrating over the 
disk of small radius $\varepsilon$ containing the string core as
\begin{eqnarray}
 \mu_I = \int_0^{\varepsilon} dr \, \sigma^2 \sqrt{\gamma} f_I(y,r)  ~,
\end{eqnarray}
where $I = 0, y, yr, r, \theta$.
Even without specifying the warp factors inside the string core explicitly, the Einstein equations 
(\ref{WS:Einsteinmunu})-(\ref{WS:Einsteintt}) 
and the boundary conditions (\ref{WS:stringboundary}) require the components of the string tension to satisfy the following relations
\begin{eqnarray}
& & \left[ \sigma ( \partial_r \sigma ) \sqrt{\gamma} \right]_0^{\varepsilon} = - \frac{2}{M_7{}^5}(\mu_r + \mu_\theta)
 ~,
\label{WS:stringtension1} \\
& & \left[ \sigma^2 \partial_r \sqrt{\gamma} \right]_0^{\varepsilon} = - \frac{1}{4 M_7{}^5} ( 4 \mu_0 + \mu_r - 3 \mu_\theta ) ~,
\label{WS:stringtension2} \\
& & 4 \mu_0 - 4 \mu_y + \mu_r + \mu_\theta = 0  ~,
\end{eqnarray}
neglecting the order ${\cal O}(\varepsilon)$ contributions.
These quantities can be obtained solely in terms of 
the values of warp factors at the boundaries of the core $r=0$ and $\varepsilon$, 
without specifying the warp factors explicitly inside the 
string core. 
Note that $\mu_I$ are functions of $y$ as there are warp factors
which change the scale of the theory along the $y$-direction.
Since our solutions (\ref{WS:metricansatzabcd}) relate the left-hand sides of 
Eqs.(\ref{WS:stringtension1}) and (\ref{WS:stringtension2}), 
we find the following relation between 
the components of the string tension, 
\begin{eqnarray}
 \pm a ( \mu_y - \mu_\theta ) + 2b  M_7{}^5 ( \mu_r + \mu_\theta ) 
&=& b \sigma^2(y,0)
 ~.
\end{eqnarray}

\begin{figure}
\begin{center}
\leavevmode
\epsfxsize=13cm
\epsfbox{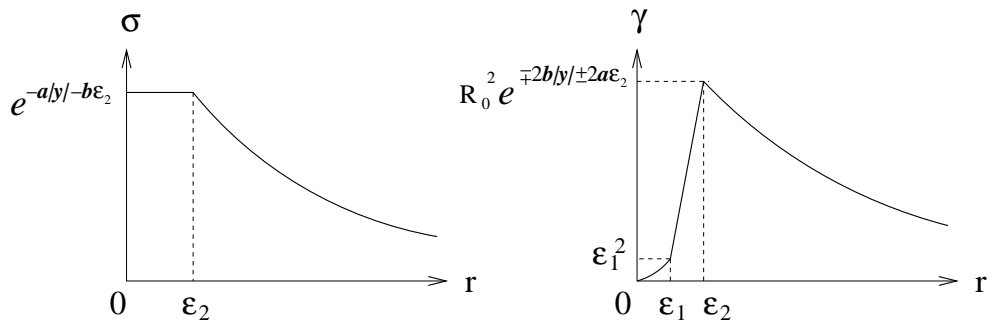}
\end{center}
\caption[]{The warp factors near the core of 
the string-like defect at $y={\rm constant}$ as a function of 
 $r$ (\ref{WS:solutionexample}). 
}
\label{WS:fig-sigma-gamma}
\end{figure}

Since more explicit form of the warp factors 
inside the core $r < \varepsilon$ of the string-like defect 
 is needed to obtain each component of the 
energy-momentum tensor $\hat T^{({\rm string})}$ separately, 
 we will take the following warp factors inside the core as an 
illustrative example 
\begin{eqnarray}
 \sigma(y,r) &\!\!\!=&\!\!\! e^{-a|y|-b \varepsilon_2} 
\Theta(\varepsilon_2 - r) + e^{-a|y|-br} \Theta(r - \varepsilon_2) 
~,  \nonumber \\
 \gamma(y,r) &\!\!\!=&\!\!\! r^2 \Theta(\varepsilon_1 - r) 
+ \left\{ A(y) (r-\varepsilon_1) + \varepsilon_1^2 \right\} 
\left\{ \Theta(r - \varepsilon_1) - \Theta(r - \varepsilon_2) \right\}
\nonumber \\
 &\!\!\! &\!\!\! {} + R_0{}^2 e^{\mp 2b|y| \pm 2ar} 
\Theta(r - \varepsilon_2) ~,
\label{WS:solutionexample}
\end{eqnarray}
where $\Theta$ is a step function, 
$0 < \varepsilon_1 < \varepsilon_2 = \varepsilon - 0$ and
\begin{eqnarray}
 A(y) \equiv \frac{R_0{}^2 e^{\mp 2b|y| \pm 2a \varepsilon_2} 
- \varepsilon_1^2}{\varepsilon_2 - \varepsilon_1} ~.
\end{eqnarray}
The behavior of the warp factors are 
 shown in Fig.\ref{WS:fig-sigma-gamma},
The components of the string tension needed for this solution of the 
warp factors are given for $y\not =0$ as 
\begin{eqnarray}
 \mu_0 &\!\!\!=&\!\!\! M_7{}^5 
\left\{ \left( \frac{3}{2}b \pm a \right) R_0 e^{-(2a \pm b)|y|} 
- e^{-2a|y|} \right\}
~, \\
\mu_y &\!\!\!=&\!\!\! M_7{}^5 
\left\{ \left( 2b \pm a \right) R_0 e^{-(2a \pm b)|y|} - e^{-2a|y|} \right\}
~, \\
 \mu_{yr} &\!\!\!=&\!\!\! \mp b R_0 M_7{}^5 |y|' e^{-(2a \pm b)|y|}
~, \label{WS:stringtensionyr}\\
 \mu_r &\!\!\!=&\!\!\! 0
~, \\
 \mu_\theta &\!\!\!=&\!\!\! 2 b R_0 M_7{}^5 e^{-(2a \pm b)|y|} ~.
\end{eqnarray}
We observe that $\mu_{yr}(y)$ has a discontinuity at the wall $y=0$. 
Since  $\mu_{yr}(y)$ is an odd function of $y$, the nonvanishing value of 
 $\mu_{yr}(y\rightarrow +0)$ implies the discontinuity. 
We believe that the discontinuity comes about because we assumed a vanishing 
width for the wall, and that it will be smoothed out if the four-brane 
and five-branes are intersecting smoothly. 
Therefore we should understand the solution 
(\ref{WS:metric1}), (\ref{WS:metricansatzabcd}), 
(\ref{WS:solutionexample}) as a limit of the 
vanishing width for the walls including the intersection region. 

When we assume the above warp factors (\ref{WS:solutionexample}) 
even at the intersection region, we also obtain the energy-momentum tensor 
for the walls 
even inside the string core $0 \le r < \varepsilon$. 
We find, for instance, $t^{(1)}$'s are modified from 
Eq.(\ref{WS:emsolution1}) by multiplying the constant $b$ with 
a function $B(r)$ 
\begin{equation}
 b \rightarrow b B(r), \qquad 
B(r)\equiv \left[1+{\varepsilon_2-r \over r-\varepsilon_1}
\left({\varepsilon_1 \over R_0}\right)^2 {\rm e}^{\mp 2a \varepsilon_2}
\right]^{-1} ~.
\end{equation}
Therefore the components of the energy-momentum tensor $t^{(1)}$'s on the wall 
change from the constant value (\ref{WS:emsolution1}) outside 
 the string core $ r > \varepsilon$ 
to another constant value at the center ($0\le r < \varepsilon_1$) smoothly  
through the string core region $0 \le r < \varepsilon$ 
where the wall intersects with the string-like defect. 
The components of the energy-momentum tensor on the other wall $t^{(2)}$'s also exhibit a similar behavior. 

The four-dimensional reduced Planck scale $M_4$ is related to the
seven-dimensional Planck scale $M_7$ as
\begin{equation}
 M_4{}^2 = M_7{}^5 \int_0^{y_c} dy \int_0^\infty dr \int_0^{2\pi} d\theta \, 
\sigma \sqrt{\gamma} ~,
\label{WS:4DPlanck}
\end{equation}
which gives after integration
\begin{equation}
 M_4{}^2 = \pm \frac{2 \pi R_0}{b^2 - a^2} 
\left( 1 - e^{-(a \pm b)y_c} \right)  M_7{}^5 
\end{equation}
for $a \pm b \ne 0$, and
\begin{equation}
 M_4{}^2 = \frac{\pi R_0 y_c}{b} M_7{}^5
\end{equation}
for $a \pm b = 0$.
If we want to consider a single  wall configuration 
 which is obtained by taking $y_c \to \infty$
and by removing the regulator brane at $y=y_c$,
we must choose a solution which satisfies $a \pm b > 0$.

The choice of $b=0$ reduces to a rather trivial solution giving a 
tensor product of the model of parallel walls in the five-dimensional 
spacetime~\cite{ph/9905221}\cite{th/9906064} and a two-dimensional model 
with a string-like defect similar to Ref.\cite{th/0004014}, because then 
the $(r,\theta)$-directions become completely separated from the other 
directions. 
Instead, in the rest of this paper, we will concentrate on the $a=0$ 
case solution and
choose the upper sign\footnote{
The change of sign merely exchanges the role of two walls in this case. 
The choice of the upper sign corresponds to choosing the wall at $y=0$ 
as the brane with positive cosmological constant. 
}
 in (\ref{WS:solution1}) in order to obtain a finite
four-dimensional reduced Planck scale even in the limit $y_c \to \infty$.
We will eventually take $y_c \to \infty$ to consider the configuration of 
a single wall intersecting with a string-like defect at $y=0$ and 
$r < \varepsilon$.
The metric solution for $r>\varepsilon$ 
(outside of the string-like defect core) 
now becomes
\begin{eqnarray}
 ds^2 = e^{-br}\eta_{\mu\nu} dx^\mu dx^\nu - dy^2 - dr^2 
- R_0{}^2 e^{-2b|y|}d\theta^2 ~,
\label{WS:metrica0}
\end{eqnarray}
where
\begin{equation}
 b = \sqrt{\frac{-2 \Lambda}{5 M_7{}^5}} ~.
\label{WS:value-of-b}
\end{equation}
Taking the upper sign in Eq.(\ref{WS:metricansatzabcd}), 
excluding the regulator wall at $y=y_c$, and setting $a=0$ in 
Eq.(\ref{WS:emsolution1}), we find the energy-momentum tensor of 
matter fields on the wall at $y=0$ 
 satisfying the Einstein equations as 
\begin{equation}
 t_0^{(1)}(r) =  
    2b M_7{}^5  - \Lambda_6^{(1)}    
~, 
\qquad 
 t_r^{(1)}(r) =  
    2b M_7{}^5  - \Lambda_6^{(1)}    
~,
\qquad 
 t_\theta^{(1)}(r) = 
   - \Lambda_6^{(1)}  ~.
\label{WS:wall-em-tensor}
\end{equation}
In this case, we can explicitly construct the energy-momentum 
tensor on the  wall at $y=0$ by means of matter fields. 
Let us consider two scalar fields  $\varphi^a, a=1,2$ on the  wall 
with a Lagrangian density~\cite{th/0006251} 
\begin{eqnarray}
 {\cal{L}}_{{\rm wall}}^{(1)} 
 &\!\!\! = &\!\!\!  \frac{1}{2} \gbar^{A B} 
\partial_{A} \varphi^{{a}} \partial_{B} \varphi^{{a}} - V(\varphi) ~,
\nonumber \\
 V(\varphi)  &\!\!\! = &\!\!\!  \lambda 
(\varphi^{{a}} \varphi^{{a}} - v^2)^2  ~,
\label{WS:scalarpotential}
\end{eqnarray}
where the indices $A, B$ span only over coordinates on the wall 
$A, B = 0, \cdots, 3, r, \theta$. 
The energy-momentum tensor is then given by 
\begin{eqnarray}
\bar{T}^{({\rm wall} 1)}{}^{A}_{B} 
= \gbar^{A C} \partial_{C} \varphi^{{a}} \partial_{B} \varphi^{{a}} 
- \delta^{A}_{B} \left( \frac{1}{2} \gbar^{C D} 
\partial_{C} \varphi^{{a}} \partial_{D} \varphi^{{a}} - V(\varphi) \right)   ~.
\nonumber
\end{eqnarray}
The minimum of the potential $V$ in Eq.(\ref{WS:scalarpotential}) is at 
$\varphi^{{a}} \varphi^{{a}} = v^2$. 
Therefore we can  have a winding configuration of the scalar fields 
\begin{equation}
 \varphi^1 = v \cos \theta  ~, \quad \varphi^2 = v \sin \theta ~,
\end{equation}
which becomes stable for large values of $\lambda$. 
Then the scalar field gives an energy-momentum tensor on 
the  wall with components 
\begin{equation}
 t_0^{(1)} = t_r^{(1)} = - t_\theta^{(1)} = \frac{v^2}{2 R_0{}^2} ~.
\end{equation}
Therefore the energy-momentum tensor (\ref{WS:wall-em-tensor}) satisfying 
the Einstein equation is obtained 
if the vacuum expectation value of the scalar fields is related to 
the brane tension $\Lambda_6^{(1)}$, the Planck scale $M_7$ 
in the bulk and the warp factor slope 
$b$ in Eq.(\ref{WS:value-of-b}) as 
\begin{equation}
 \Lambda_6^{(1)} = M_7{}^5 b = \frac{v^2}{2 R_0^2} ~.
\end{equation}

\section{Graviton on the Background of a Wall and a String}
\subsection{Massless and Massive Spin-2 Fields}

Having obtained a solution for the system, we wish to find the modes 
on the background of the solution. 
We first examine whether spin 2 graviton is localized on the
four-dimensional intersection region, and then check whether 
the resulting gravitational force is 
consistent with the four-dimensional experimental gravity.
{}For these purposes, let us consider linearized equations for 
 gravitons. 
We will only consider the spin 2 fields and neglect the spin 1 and spin 0 
fields, which must be included for the complete analysis eventually.
Let us consider the following fluctuations of metric,
\begin{equation}
 \ghat_{\mu \nu} = \sigma(r) \left[ \eta_{\mu \nu} 
+ h_{\mu \nu}(x^A) \right] ~.
\label{WS:metricfluctuation}
\end{equation}
We will choose a gauge \cite{th/9906064}
\begin{equation}
\partial_\mu h^{\mu}{}_\nu (x^A)=h^\mu{}_\mu (x^A)=0 ~. 
\end{equation}
Then the linearized equations of motion takes the form~\cite{th/0006203}
\begin{equation}
 \frac{1}{\sqrt{\ghat}} \partial_A 
\left( \sqrt{\ghat} \ghat^{AB} \partial_B h_{\mu \nu}(x^A) \right) = 0 ~.
\label{WS:linearizedeq2}
\end{equation}
Since we take $a=0$ case (\ref{WS:metrica0}), we can separate the variables as
\begin{eqnarray}
 h_{\mu\nu}(x^A) = \sum_{m,k,\ell} h^{(mk\ell)}_{\mu\nu}(x^\alpha) 
\phi_{(mk)}(r) \chi_{(k \ell)}(y) e^{i \ell \theta} ~,
\label{WS:decomposition1}
\end{eqnarray}
where $\ell$ is an integer angular momentum.
Substituting the decomposition (\ref{WS:decomposition1}) 
into the linearized equations of motion~(\ref{WS:linearizedeq2}), 
we obtain the following separated equations 
\begin{eqnarray}
 - \eta^{\lambda \rho} \partial_\lambda \partial_\rho h^{(mk\ell)}_{\mu \nu}
 &\!\!\!=&\!\!\!  m^2 h^{(mkl)}_{\mu \nu}  ~,
\label{WS:phieq1} \\
 - \partial_r \left( \sigma^2 \partial_r \phi_{(mk)} \right) 
+ k^2 \sigma^2 \phi_{(mk)} &\!\!\!=&\!\!\! m^2 \sigma \phi_{(mk)}  ~,
\label{WS:phieq2} \\
 - \partial_y \left( \gamma^{\frac{1}{2}} \partial_y \chi_{(k \ell)} \right) 
+ \ell^2 \gamma^{-\frac{1}{2}} \chi_{(k \ell)} 
&\!\!\!=&\!\!\! k^2 \gamma^{\frac{1}{2}} \chi_{(k \ell)}  ~.
\label{WS:chieq2}
\end{eqnarray}
Eq.(\ref{WS:phieq1}) shows that the eigenvalue $m \ge 0$ corresponds to the 
mass of the four-dimensional field $h^{(mkl)}_{\mu \nu}(x^\alpha)$.
{}For the differential operator in the left-hand side of Eq.(\ref{WS:chieq2}) 
to be self-adjoint we impose the boundary conditions\footnote{
Another possible boundary condition is the Dirichlet boundary condition 
$\chi_{(k \ell)}(0) = \chi_{(k \ell)}(y_c) = 0$. 
However, modes satisfying the Dirichlet boundary condition do not contain 
the zero mode that we are most interested in. 
Moreover $Z_2$ projection allows only the Neumann boundary condition (\ref{WS:selfadjointchi}).
To obtain the corrections to the Newton potential 
(\ref{WS:nonzero-mode-potential}), 
it is sufficient to consider 
(\ref{WS:selfadjointchi}).
}
\begin{equation}
 \chi_{(k \ell)}'(0) = \chi_{(k \ell)}'(y_c) = 0  ~.
\label{WS:selfadjointchi}
\end{equation}
Then, we can demand the modes $\chi_{(k \ell)}$ to satisfy the 
orthonormal condition
\begin{equation}
 \int_0^{y_c} dy \, \gamma^{\frac{1}{2}} \chi_{(k \ell)}^* \chi_{(k' \ell)} 
= \delta_{k k'}  ~.
\label{WS:orthonormalchi}
\end{equation}
Multiplying $\chi_{(k \ell)}$ to Eq.(\ref{WS:chieq2}) and integrating 
the first term of the left-hand side by parts, we obtain 
\begin{equation}
 k^2 = \int_0^{y_c} dy \, \gamma^{\frac{1}{2}} 
\left| \chi_{(k \ell)}' \right|^2 
+ \ell^2 \int_0^{y_c} dy \, \gamma^{- \frac{1}{2}} 
\left| \chi_{(k \ell)} \right|^2
 \ge 0  ~.
\label{WS:ksquare}
\end{equation}
This equation together with Eq.(\ref{WS:orthonormalchi}) shows that $k^2$ is 
non-negative, more precisely 
\begin{eqnarray}
 \ell \ne 0 \,\,&\Longrightarrow& k^2 > \frac{1}{R_0{}^2} ~,
\label{WS:gap1}
\\
 \ell=0 \,\,&\Longrightarrow& k^2 \ge 0 ~, 
\label{WS:gap2}
\end{eqnarray}
where the equality $k^2=0$ holds if and only if $\chi_{(00)}(y)$ 
is independent of $y$.
Our boundary condition (\ref{WS:selfadjointchi}) is the most natural one as 
long as the gravity can be treated semiclassically. 
We will examine the question of the validity of semiclassical gravity 
at the end of this subsection. 

A similar analysis applies to Eq.(\ref{WS:phieq2}).
We can impose the boundary conditions\footnote{
We do not choose the Dirichlet boundary condition by the same reason as that 
for $\chi_{(kl)} (y)$ except that we do not have $Z_2$ projection in $r$. 
}
\begin{equation}
 \dot{\phi}_{(mk)}(0) = \dot{\phi}_{(mk)}(\infty) = 0  ~,
\label{WS:selfadjointphi}
\end{equation}
for the differential operator in Eq.(\ref{WS:phieq2}) to be self-adjoint.
Then the modes $\phi_{(mk)}$ can be made to satisfy the orthonormal condition
\begin{equation}
 \int_0^\infty dr \, \sigma \phi_{(mk)}^* \phi_{(m'k)} = \delta_{m m'}  ~.
\label{WS:orthonormalphi}
\end{equation}
where quantum numbers $k$ are not summed. 
Multiplying $\phi_{(mk)}^*$ to Eq.(\ref{WS:phieq2}) and integrating the first
term of the left-hand side by parts, we obtain
\begin{equation}
 m^2 = \int_0^\infty dr \, \sigma^2 
\left| \dot{\phi}_{(mk)} \right|^2 
+ k^2 \int_0^\infty dr \, \sigma^2 \left| \phi_{(mk)} \right|^2
 \ge 0 ~,
\label{WS:msquare}
\end{equation}
where the equality $m^2=0$ holds if and only if $k=0$ and $\phi_{(00)}(r)$ 
is independent of $r$. 
We now obtain from Eqs.(\ref{WS:ksquare}) and (\ref{WS:msquare}) that $m=0$ 
requires $k=\ell=0$. 
Since $m$ is the mass of the four-dimensional field 
$h^{(mk\ell)}_{\mu\nu}(x^\alpha)$, 
we find that there is only one massless four-dimensional mode for spin 2 
graviton as required by experiment. 

It is easy to see from the wave equations~(\ref{WS:phieq2}),
(\ref{WS:chieq2}) and the orthonormal conditions~(\ref{WS:orthonormalchi}), 
(\ref{WS:orthonormalphi}) that the unique massless mode is the zero-mode
solution with $m = k = \ell = 0$, 
\begin{eqnarray}
 \chi_{(00)}(y) = \sqrt{\frac{b}{R_0(1-{\rm e}^{-by_c})}} ~, 
\qquad \phi_{(00)}(r) = \sqrt{b} ~.
\label{WS:zeromodesolution}
\end{eqnarray}
These constant wave functions are normalizable. 
The definition (\ref{WS:4DPlanck}) of the four-dimensional reduced 
Planck scale implies that 
the probability density of the wave functions 
$\phi_{(mk)}(r) \chi_{(k \ell)}(y) $ in the extra dimensions is given by 
$\sigma(r) \gamma^{1\over 2}(y) |\phi_{(mk)}(r) \chi_{(k \ell)}(y)|^2$. 
The probability density of the zero mode $m=k=\ell=0$  wave function 
is sharply peaked around 
 the intersection $y=0, r=0$. 
Therefore the spin 2 graviton represented by the zero mode is 
localized around the intersection region. 

In order to evaluate the density of states of the non-zero modes 
 precisely, we reintroduce a regulator brane at 
a finite radial distance cutoff $r_{\rm max}$. 
Accordingly, the boundary condition 
(\ref{WS:selfadjointphi}) at $r=\infty$ is modified to a similar boundary 
condition imposed at $r=r_{\rm max}$. 

Now consider the non-zero modes.
{}First, let us solve the wave equation (\ref{WS:chieq2}) for $\chi_{(k \ell)}(y)$.
\begin{flushleft}
 (i) $\ell = 0$ modes
\end{flushleft}
The $k=0$ mode solution is a constant $\chi_{(00)}$ given in 
Eq.(\ref{WS:zeromodesolution}).
The $0 < k \le \frac{b}{2}$ mode solutions turn out to be excluded from
the spectrum by the boundary conditions (\ref{WS:selfadjointphi}).
The $k>\frac{b}{2}$ mode solutions are 
\begin{equation}
 \chi_{(k0)} = N_{k0} e^{\frac{b}{2}|y|} \left( e^{-i \frac{1}{2} \sqrt{4k^2-b^2}|y|} + \alpha_{k0} e^{i \frac{1}{2} \sqrt{4k^2-b^2}|y|} \right)  ~,
\label{WS:solutionchi0k}
\end{equation}
where $N_{k0}$ are normalization constants and $\alpha_{k0}$ are constants
which are determined by the boundary condition at $y=0$ as
\begin{equation}
 \alpha_{k0} = - \frac{b - i \sqrt{4k^2-b^2}}{b + i \sqrt{4k^2-b^2}} ~.
\end{equation}
In addition, there is a boundary condition at $y=y_c$ which leads to the
quantization of $k$ as
\begin{equation}
 k^2 = \frac{b^2}{4} + \frac{\pi^2}{y_c^2} n_k^2   ~,
\end{equation}
where $n_k =1,2,\cdots$.
Thus we find that there is a gap in the spectrum of $k$ between 
$k=0$ and $k > b/2$ even in the limit of $y_c \rightarrow \infty$.
\begin{flushleft}
 (ii) $\ell \ne 0$ modes
\end{flushleft}
Changing the variable from $y$ to $u \equiv \frac{\ell}{b R_0} e^{b|y|}$ 
and redefining the wave equation by 
$\tilde{\chi}_{(k \ell)} \equiv u^{-\frac{1}{2}} \chi_{(k \ell)}$, 
Eq.(\ref{WS:chieq2}) can be rewritten in the form
\begin{eqnarray}
 \left[ \frac{d^2}{du^2}  + \frac{1}{u} \frac{d}{du} 
- \left\{ 1 + \frac{1}{u^2} 
\left( \frac{1}{4} - \frac{k^2}{b^2} \right) \right\} \right] 
\tilde{\chi}_{(k \ell)} = 0  ~.
\end{eqnarray}
The solution to this equation is a linear combination of modified Bessel 
 functions.
Therefore, the solution is of the form
\begin{eqnarray}
 \chi_{(k \ell)} = N_{k \ell} u^\frac{1}{2} 
\left[ K_{\nu_k}(u) + \alpha_{k \ell} I_{\nu_k}(u) \right] ~,
\end{eqnarray}
where
\begin{eqnarray}
 \nu_k = i \sqrt{\frac{k^2}{b^2} - \frac{1}{4}}  ~,
\end{eqnarray}
 $N_{k \ell}$ are normalization constants and $\alpha_{k \ell}$ are constant
	   coefficients.
Irrespective of the detailed spectrum for the $\ell \ne 0$ modes,  
Eq.(\ref{WS:gap1}) shows that there is a gap in the spectrum of $k$ 
as $k>1/R_0$.

Next, let us turn to Eq.(\ref{WS:phieq2}) to obtain the wave function 
$\phi_{(mk)}(r)$.
\begin{flushleft}
 (i') $m = 0$ mode
\end{flushleft}
As stated before, there is only one $m=0$ mode corresponding to $m=k=\ell=0$.
The $m=0$ mode solution is a constant which reduces to the $\phi_{(00)}$ in 
Eq.(\ref{WS:zeromodesolution}) in the limit $r_{\rm max} \to \infty$.
\begin{flushleft}
 (ii') $m > 0$ modes
\end{flushleft}
Changing the variables to $v \equiv \frac{2m}{b} e^{\frac{b}{2} r}$ and
 redefining the wave equations by $\tilde{\phi}_{(mk)} \equiv v^{-2}
 \phi_{(mk)}$, Eq.(\ref{WS:phieq2}) can be rewritten in the form
\begin{eqnarray}
 \left[ \frac{d}{d v^2} + \frac{1}{v} \frac{d}{d v} 
\left\{ 1 - \frac{4}{u_m^2} \left( 1 + \frac{k^2}{b^2} \right) \right\} 
\right] \tilde{\phi}_{(mk)} = 0  ~.
\end{eqnarray}
The solutions to this equation are linear combinations of Bessel functions.
Therefore the solution takes the form
\begin{eqnarray}
 \phi_{(mk)} = \tilde{N}_{mk} e^{br} \left[ J_{\tilde{\nu}_k}(v) 
+ \beta_{mk} Y_{\tilde{\nu}_k}(v) \right]  ~,
\end{eqnarray}
where $\tilde{N}_{mk}$ are the normalization constants, $\beta_{mk}$ are the
 constant coefficients and
\begin{eqnarray}
 \tilde{\nu}_k = 2 \sqrt{1 + \frac{k^2}{b^2}}  ~.
\end{eqnarray}
{}From the boundary condition (\ref{WS:selfadjointphi}) at $r=0$, we obtain
\begin{eqnarray}
 \beta_{mk}
 = - \frac{ (2-\tilde{\nu}_k) J_{\tilde{\nu}_k}(v_0) 
+ v_0 J_{\tilde{\nu}_k - 1}(v_0) }
{ (2-\tilde{\nu}_k) Y_{\tilde{\nu}_k}(v_0) 
+ v_0 Y_{\tilde{\nu}_k - 1}(v_0)} ~,
\label{WS:betar0}
\end{eqnarray}
where $v_0 \equiv \frac{2m}{b}$ and 
$\tilde{\nu}_k = 2 \sqrt{1+\frac{k^2}{b^2}}$ .
Since the boundary condition at $r=r_{\rm max}$ gives 
$\dot{\phi}_{(m k)} (r_{\rm max})= 0$, we find 
\begin{eqnarray}
 \beta_{mk}
 &\!\!\!=&\!\!\! - \frac{(2-\tilde{\nu}_k) J_{\tilde{\nu}_k}(v_{\rm max}) 
+ v_{\rm max} J_{\tilde{\nu}_k - 1}(v_{\rm max})}
{ (2-\tilde{\nu}_k) Y_{\tilde{\nu}_k}(v_{\rm max}) 
+ v_{\rm max} Y_{\tilde{\nu}_k - 1}(v_{\rm max})} 
\nonumber \\
 &\!\!\!\simeq&\!\!\! - \frac{J_{\tilde{\nu}_k - 1}(v_{\rm max})}
{ Y_{\tilde{\nu}_k - 1}(v_{\rm max})} 
~,
\label{WS:betarmax}
\end{eqnarray}
where $v_{\rm max} \equiv \frac{2m}{b} e^{\frac{b}{2} r_{\rm max}}$.
In the limit $\frac{m}{b} \ll 1$, Eq.(\ref{WS:betar0}) for the $k=0$ 
mode is approximated by
\begin{eqnarray}
 \beta_{m0} = - \frac{J_1(v_0)}{Y_1(v_0)} \simeq \frac{\pi}{4} v_0^2 \ll 1 ~.
\label{WS:betam0}
\end{eqnarray}
Eqs.(\ref{WS:betarmax}) and (\ref{WS:betam0}) imply that the mass spectrum is
quantized as
\begin{eqnarray}
 m \equiv m_n \simeq \left( n + \frac{1}{4} \right) \frac{\pi b}{2} 
e^{-\frac{b}{2} r_{\rm max}} ~,
\label{WS:massquantization}
\end{eqnarray}
where $n$ is an integer. The mode function $\phi_{(m0)}$ at $r=0$ becomes
\begin{eqnarray}
 \left| \phi_{(m0)}(0) \right|^2 \simeq \left| \tilde{N}_{m0} \right|^2 
 \simeq \pi m e^{-\frac{b}{2}r_{\rm max}} ~.
\label{WS:phi0valuek0}
\end{eqnarray}
On the other hand, Eq.(\ref{WS:betar0}) for the $k>0$ modes in the limit
$\frac{m}{b} \ll 1$ is approximated by
\begin{eqnarray}
 \beta_{mk} \sim v_0^{2 \tilde{\nu}_k} \ll 1 ~.
\end{eqnarray}
So the mass spectrum is quantized as
\begin{eqnarray}
 m \equiv m_n \simeq \left( n + \frac{\tilde{\nu}_k}{2} 
- \frac{3}{4} \right) \frac{\pi b}{2} e^{-\frac{b}{2} r_{\rm max}} ~,
\label{WS:massquantizationk}
\end{eqnarray}
and the mode function $\phi_{(mk)}$ at $r=0$ becomes 
\begin{eqnarray}
 \left| \phi_{(mk)}(0) \right|^2 \sim \left| \tilde{N}_{mk} \right|^2 
v_0^{2 \tilde{\nu}_k} ~.
\label{WS:phi0valuekp}
\end{eqnarray}

Let us recall that there is a gap in the spectrum of $k$ 
between $k=0$ and $k>b/2$ for the $\ell=0$ modes and $k > 1/R_0$ 
for the $\ell\not=0$ modes. 
On the other hand, the normalization factor $\tilde{N}_{mk}$ is insensitive 
to the values of $k$. 
Therefore we find that 
\begin{eqnarray}
 \left| \phi_{(m0)}(0) \right|^2 \gg \left| \phi_{(mk)}(0) \right|^2 ~,
\end{eqnarray}
and that the wave function at the intersection $r=0$ is negligible for 
the modes with nonvanishing values of $k$.

Let us discuss the question of the validity of semiclassical gravity. 
The metric (\ref{WS:metrica0}) is locally a product of $AdS_6$ 
(for $r, x^\mu$) and Euclidean $AdS_2^E$ (for $y, \theta$) 
and exhibits a conical singularity near 
$y \rightarrow \infty$, since the effective radius for the angular 
variable $\theta$ shrinks to zero. 
Because of this conical singularity, semiclassical gravity may not be 
reliable as the regulator brane goes to infnity $y_c \rightarrow \infty$. 
Therefore our boundary condition (\ref{WS:selfadjointchi}) 
at $y=y_c$ is, strictly speaking, reliable only for the regulator brane 
not too much far away from our brane and 
may be modified as $y_c \rightarrow \infty$. 

Recently a nice resolution of a conical singularity 
has been proposed by Ponton and Poppitz \cite{PP} in a 
simpler case of string-like defect in six-dimensional gravity. 
They observed that the string theory as the fundamental theory of 
gravity offers a dual description of the strong gravity in terms of the 
conformal field theory. 
Starting from a solution of supergravity corresponding to the type $I'$ 
D4-D8 brane system, they obtained the $AdS_6 \times S^4$ gravity background. 
{}From the AdS/CFT correspondence and a chain of duality arguments, 
they concluded that one way of resolving the singularity leads to a 
semiclassical gravity description for a long-distance behavior 
of the gravitational field which is the same 
as the Randall-Sundrum model in five dimensions. 
Another resolution provides a mass gap for all the Kaluza-Klein modes. 
Consequently the gravitational Newton potential for the 
first case receives the same power corrections 
as those of the model with one extra dimension 
(i.e.~the Randall-Sundrum model \cite{th/9906064}), 
whereas the gravitational Newton potential for the second case 
receives only the exponentially suppressed corrections 
corresponding to no extra dimensions effectively. 

In order to address the issue of strong gravity and the resulting possible 
modification of the boundary condition (\ref{WS:selfadjointchi}) at $y=y_c$ 
for our model, we need to find a system of branes in string theories 
which provides a solution of supergravity corresponding 
locally to our metric (\ref{WS:metrica0}) in the Maldacena limit. 
We expect that the solution corresponds to a number of coincident 
branes localized on other branes, in order to provide two radial coordinates 
$r$ for $AdS_6$ and $y$ for $AdS_2^E$. 
Semi-localized solutions have been obtained for cases which reduce 
to a product of an $AdS$ space and compact internal spaces such as 
spheres \cite{CLPV}. 
It is an extremely interesting problem to work out those supergravity 
solutions which give a product of noncompact ``internal space'' and 
the $AdS$ space and to investigate the corresponding system of branes in 
string theories. 
We hope to address these problems in future. 
The result of ref.\cite{PP} suggests that 
there may be choices of boundary conditions also in our case. 
On the other hand, our analysis using the semiclassical gravity gives 
a mass gap which makes two of the extra dimensions ineffective at 
low energies. 
This will lead to the correction of the Newton 
potential with the same power as the model with only one extra dimension 
(i.e.~the Randall-Sundrum model \cite{th/9906064}), 
 as we will see in the next subsection. 
It is interesting to note that this result resembles the result 
of ref.\cite{PP} which takes full account of the strong gravity effects.

\subsection{Gravitational potential}

If we place point sources $\bar m_1$ and $\bar m_2$ on the 
four-dimensional intersection ($r=y=0$) with a large distance 
$R\equiv \sqrt{(x^1)^2+(x^2)^2+(x^3)^2}$ apart, 
we obtain a gravitational potential coming from the exchanges of zero mode and 
nonzero modes. 
The zero mode for spin 2 graviton gives the usual Newton's law 
\begin{eqnarray}
  V(R) = - G_N \frac{\bar{m}_1 \bar{m}_2}{R}, \qquad 
G_N \equiv {1 \over 8\pi M_4^2} ~, 
\label{WS:newtonpotential}
\end{eqnarray}
where $G_N$ is the Newton's constant and $M_4$ is  the reduced Planck scale 
defined in Eq.(\ref{WS:4DPlanck}).
The nonzero modes contribute corrections to the Newton potential 
as \cite{BrandhuberSfetsos}
\begin{eqnarray}
 \Delta V(R) = - G_N \frac{\bar{m}_1 \bar{m}_2}{R} 
\sum_n \sum_{k} \sum_{l}  
\left| \frac{\phi_{(mk)}(0) \chi_{(k\ell)}(0)}
{\phi_{(00)}(0) \chi_{(00)}(0)} \right|^2 e^{-m_n R} ~.
\label{WS:nonzero-mode-potential}
\end{eqnarray}
Because of the gap in the spectrum of $k$ and $m$, 
contributions from the $k >0$ modes 
are suppressed exponentially. 
The leading contribution to the gravitational potential 
 comes from the exchange of the nonzero modes of $m \ge 0, k=\ell=0$. 
In the limit $r_{\rm max} \to \infty$, the 
spectrum~(\ref{WS:massquantization}) becomes continuous and the 
summation of $n$ is converted into an integral
\begin{eqnarray}
 \sum_n \simeq \frac{2}{\pi b} e^{\frac{b}{2} r_{\rm max}} \int_0^\infty dm ~.
\end{eqnarray}
Consequently, Eq.(\ref{WS:nonzero-mode-potential}) becomes
\begin{eqnarray}
 \Delta V(R) 
  &\simeq& - G_N \frac{\bar{m}_1 \bar{m}_2}{R} \frac{2}{b^2} 
\int_0^\infty dm \, m e^{-m R}
\nonumber \\
  &=& - \frac{2 G_N}{b^2} \frac{\bar{m}_1 \bar{m}_2}{R^3} ~.
\label{WS:correctionpotential2}
\end{eqnarray}

One should note that 
the correction term behaves $\Delta V(R) \sim 1/R^3$ in the case of 
the wall in five dimensions~\cite{th/9906064}, and $\Delta V(R) \sim 1/R^4$
in the case of the string-like defect in six 
dimensions~\cite{th/0004014}\footnote{
This $1/R^4$ behavior of the correction term is modified to $1/R^3$ 
behavior or to no power corrections 
by taking acount of the strong gravity effects in ref.\cite{PP}.
}. 
This is natural since the gravitational flux tends to spread out if 
there are more extra dimensions. 
The correction term (\ref{WS:correctionpotential2}) to the Newton potential 
 $\Delta V(R) \sim 1/R^3$ 
 falls off precisely like the case of a  wall in five 
dimensions even though we have seven dimensional bulk spacetime. 
Our counter-intuitive result can be understood as follows. 
The gap for $k >0$ modes implies that only the $k=0$ 
mode contributes without being suppressed exponentially. 
Both the gapless continuum massive graviton modes ($k=\ell=0$) as well as 
the zero mode are localized on the wall at 
$y=0$, and moreover the zero-mode ($m=k=\ell=0$) is localized at the
intersection of the wall and the string.
Since the $k=0$ mode is localized at $y=0$, the gravitational flux 
does not spread out into the direction $y$. 
This effectively reduces one dimension. 
Since the $\ell \not=0$ modes are also exponentially suppressed 
because of the mass gap, 
only the $\ell=0$ mode contributes to the leading correction. 
Therefore the angular direction is not effective either. 
On the other hand, the remaining direction $r$ acts precisely like 
the coordinate of the single extra dimension $y$ in the original wall model in 
five-dimensional spacetime. 
Therefore the same mechanism applies to our case as the wall model 
in five dimensions, leading to the same correction 
(\ref{WS:correctionpotential2}) as the wall model. 

We have to examine the contributions from spin 1 and spin 0 fields 
coming from the seven-dimensional graviton in order to obtain the full 
gravitational potential and also to ascertain the stability of the 
background metric \cite{th/9912160}, \cite{th/0010208}. 
It has also been noted that the contributions from the bending of the brane 
due to the backreaction of the matter source will modify the coefficient of 
the $1/R$ term in Eq.(\ref{WS:newtonpotential}) and consequently the relative 
normalization of the correction term (\ref{WS:nonzero-mode-potential}) 
\cite{th/9911055}. 
{}Finally we should have in mind that the strong gravity effects near 
the conical singularity as 
the regulator brane goes to infinity $y_c \rightarrow \infty$ 
may modify the boundary condition at $y=y_c$ and the corrections to 
the gravitational Newton potential. 

\section{Intersection of $n$ Walls 
and a String-like Defect}
\label{WS:INTERSECTIONn}

In the following, we extend the idea of the previous section to a 
configuration of mutually intersecting $n$  walls 
(namely $(5+n)$-branes) and a string-like defect (namely $(4+n)$-brane) 
in $(6+n)$ spacetime dimensions. 
Our four-dimensional world will come out as an intersection of these walls and 
the string-like defect. 

We can perform the extension along the lines of~\cite{th/9907209}, where 
they worked $n$ orthogonally intersecting walls in a locally $AdS_{4+n}$ 
bulk geometry. 
Since the bulk geometry of the metric~(\ref{WS:metrica0}) is locally 
$AdS_{5} \times AdS^{\rm E}_{2}$ where $AdS^{\rm E}_2$ denotes two-dimensional
Euclidean anti-de Sitter space in ($y, \theta$) directions, 
we are prompted to extend the $AdS^{\rm E}_{2}$ geometry to 
$AdS^{\rm E}_{n+1}$.
We will work on a $D \equiv 6+n$ dimensional spacetime with the coordinates 
$x^A \equiv (x^\mu,z_j,r,\theta), \; j=1, 2, \cdots, n$. 
As in the previous case, $(r,\theta)$ are
polar coordinates of extra dimensions transverse to the string-like defect, 
which is placed at $r=0$.
The coordinate for the direction 
normal to the $j$-th wall is denoted as 
$z_j$ ($j=1,2,\cdots,n$). 
The action of the extended system is
\begin{eqnarray}
 S = S_{\rm gravity} + S_{\rm string} + \sum_{j=1}^n S^{(j)}_{\rm wall} ~,
\end{eqnarray}
\begin{eqnarray}
 S_{\rm gravity} &\!\!\!=&\!\!\! 
\int d^Dx \, \sqrt{\ghat} 
\left( -\frac{1}{2}M_D{}^{D-2} \Rhat - \Lambda_D \right)
~, \\
 S_{\rm string} &\!\!\!=&\!\!\! 
\int_{r < \varepsilon} d^Dx \, \sqrt{\ghat} {\cal L}_{\rm string}
~, \\
 S^{(j)}_{\rm wall} &\!\!\!=&\!\!\! 
\int_{z_j=0} d^{D-1}x \, \sqrt{\bar{g}^{(j)}}
 {\cal L}^{(j)}_{\rm wall} ~,
\end{eqnarray}
where $M_D$ is the fundamental $D$-dimensional Planck scale and
$\Lambda_D$ is a bulk cosmological constant.
We included the cosmological constants on branes in the Lagrangians
${\cal L}_{\rm string}$ and ${\cal L}^{(j)}_{\rm wall}$'s.
The $j$-th  wall is placed at $z_j=0$, on which the induced metric is
\begin{eqnarray}
 \bar{g}^{(j)}_{A_j B_j} \equiv \ghat_{A_j B_j}|_{z_j=0} ~,
\end{eqnarray}
where $A_j, B_j$ are the $(D-1)$-dimensional indices along the $j$-th  
wall. 
Therefore the subscript $j$ on $A,B$ denotes an exclusion of the 
coordinate $z_j$.

Einstein equations for this action are
\begin{eqnarray}
 \Rhat_{AB} - \frac{1}{2}\ghat_{AB}\Rhat
 = \frac{1}{M_D{}^{D-2}} \Bigg[ \Lambda_D \ghat_{AB} 
+ \hat{T}^{({\rm string})}_{AB}
   + \sum_{j=1}^n \frac{ \sqrt{\bar{g}^{(j)}} }{ \sqrt{\ghat} } 
\tilde{T}^{(j)}_{A_j B_j} \delta^{A_j}_A \delta^{B_j}_B \delta(z_j) \Bigg] ~,
\label{WS:einstein-n}
\end{eqnarray}
where $\hat{T}^{({\rm string})}_{AB}$ and $\tilde{T}^{(j)}_{A_j B_j}$'s 
are the energy-momentum tensor obtained from ${\cal L}_{string}$ and 
${\cal L}^{(j)}_{\rm wall}$'s as 
defined in Eq.(\ref{WS:enrgy-momentum}). 
The tilde ($\, \tilde{} \,$) on  $\tilde{T}^{(j)}_{A_j B_j}$ is to denote that 
 the wall tensions are included into the energy-momentum tensor. 

We assume an ansatz for the metric of a diagonalized form with 
warp factors\footnote{
The warp factor $\Omega$ for the angular coordinate $\theta$ has to vanish 
at the origin of the two-dimensional polar coordinates $r=0$ like 
Eq.(\ref{WS:stringboundary}) in order to 
have no singularity. 
Since there is no such requirement for the other coordinates $z_j$'s, 
the warp factors $\Omega$ for 
$\theta$ and $z_j$'s can differ inside the string core 
$0 \le r < \varepsilon$. 
}
 $\sigma(r, z)$ and $\Omega(r, z)$ 
\begin{eqnarray}
 ds^2 = \sigma(r,z)\eta_{\mu\nu} dx^\mu dx^\nu - dr^2 
- \Omega^2(r,z)\left( \sum_{j=1}^n dz_j^2 + R_0{}^2 d\theta^2 \right) ~.
\label{WS:metricansatzn}
\end{eqnarray}
The ratios of the determinant of metrics 
reduce as follows,
\begin{eqnarray}
 \frac{\sqrt{\bar{g}^{(j)}}}{\sqrt{\ghat}}\delta(z_j) 
= \sqrt{-\ghat^{jj}}\delta(z_j) = \frac{1}{\Omega}\delta(z_j) ~.
\end{eqnarray}
As a natural extension of the solution considered in the last section, 
we look for a solution of the form 
\begin{eqnarray}
 \sigma(r, z) \equiv e^{-br} ~, \quad
 \Omega(r, z) \equiv \frac{1}{K \sum_{j=1}^n |z_j| + 1} ~,
\label{WS:metric-n}
\end{eqnarray}
for outside of the string-like defect core ($r>\varepsilon$).
We assume that the non-zero components of the energy-momentum tensor from the 
matter sources on the $j$-th  wall for $r>\varepsilon$ are 
\begin{eqnarray}
 \tilde{T}^{(j)}{}^\mu_\nu = \tilde{t}^{(j)}_0 \delta^\mu_\nu
~, \quad
 \tilde{T}^{(j)}{}^r_r = \tilde{t}^{(j)}_r 
~, \quad
 \tilde{T}^{(j)}{}^i_k = \tilde{t}^{(j)}_z \delta^i_k 
~, \quad
 \tilde{T}^{(j)}{}^\theta_\theta = \tilde{t}^{(j)}_\theta
~,
\label{WS:energy-momentum}
\end{eqnarray}
where $i,k \ne j$. 
We assume that the string-like defect is a strictly local defect, {\it i.e.}, 
$T^{({\rm string})}_{AB} \ne 0$ for $r < \varepsilon$ and $T^{({\rm
string})}_{AB}=0$ for $r>\varepsilon$.
{}From the metric ansatz (\ref{WS:metricansatzn}) and the energy-momentum 
tensor ansatz (\ref{WS:energy-momentum}), Einstein equations 
(\ref{WS:einstein-n}) for $r>\varepsilon$ become 
\begin{eqnarray}
 \frac{3}{2}\frac{\ddot{\sigma}}{\sigma} + \frac{n^2(n+1)}{2}K^2 
- 2n \frac{K}{\Omega}\sum_{i=1}^n \delta(z_i)
= -\frac{1}{M_D{}^{D-2}} \left[ \Lambda_D + \frac{1}{\Omega} 
\sum_{i=1}^n \tilde{t}^{(i)}_0 \delta(z_i) \right]
~,
\end{eqnarray}
\begin{eqnarray}
 \frac{3}{2}\frac{\dot{\sigma}^2}{\sigma^2} + \frac{n^2(n+1)}{2}K^2 
- 2n \frac{K}{\Omega}\sum_{i=1}^n \delta(z_i)
 = -\frac{1}{M_D{}^{D-2}} \left[ \Lambda_D 
+ \frac{1}{\Omega} \sum_{i=1}^n \tilde{t}^{(i)}_r \delta(z_i) \right]
~,
\end{eqnarray}
\begin{eqnarray}
 2\frac{\ddot{\sigma}}{\sigma} + \frac{1}{2}\frac{\dot{\sigma}^2}{\sigma^2} 
+ \frac{n^2(n-1)}{2}K^2 - 2(n-1)\frac{K}{\Omega} 
\left\{ \sum_{i=1}^n \delta(z_i) - \delta(z_j) \right\}
\nonumber\\
 = -\frac{1}{M_D{}^{D-2}} \left[ \Lambda_D 
+ \frac{1}{\Omega} \sum_{i=1}^n \left\{ \tilde{t}^{(i)}_z \delta(z_i) 
- \tilde{t}^{(j)}_z \delta(z_j) \right\} \right]
~,
\end{eqnarray}
\begin{eqnarray}
 2\frac{\ddot{\sigma}}{\sigma} + \frac{1}{2}\frac{\dot{\sigma}^2}{\sigma^2} 
+ \frac{n^2(n-1)}{2}K^2 - 2(n-1)\frac{K}{\Omega} 
\left\{ \sum_{i=1}^n \delta(z_i) - \delta(z_j) \right\}
\nonumber\\
= -\frac{1}{M_D{}^{D-2}} \left[ \Lambda_D 
+ \frac{1}{\Omega}\sum_{i=1}^n \tilde{t}^{(i)}_\theta \delta(z_i) \right] ~.
\end{eqnarray}
The solution to the above Einstein equations is found to be 
\begin{eqnarray}
 b = nK ~, \quad
 K = \sqrt{\frac{-2\Lambda_D}{n^2(n+4)M_D{}^{D-2}}} ~.
\label{WS:solutionnbK}
\end{eqnarray}
Therefore the negative cosmological constant $\Lambda_D<0$ in the bulk 
is required.
The $\delta$-functions at $z_j=0$ ($j=1,2,\cdots,n$) imply that 
the energy-momentum tensor on the walls must satisfy
\begin{eqnarray}
 \tilde{t}^{(j)}_0 = \tilde{t}^{(j)}_r = 2nK M_D{}^{D-2}
~, \quad
 \tilde{t}^{(j)}_z = \tilde{t}^{(j)}_{\theta} = 2(n-1)K M_D{}^{D-2} ~.
\end{eqnarray}
The solution inside the string-like defect core $r<\varepsilon$ can be 
obtained by adjusting appropriate string tension components 
to satisfy the boundary conditions imposed at $r=0$ as in 
Eq.(\ref{WS:stringboundary}). 
A concrete example may be worked out following the illustrative example 
in Eq.(\ref{WS:solutionexample}). 

The four-dimensional reduced Planck scale for this solution is given by 
\begin{eqnarray}
 M_4{}^2 
&\!\!\!=&\!\!\! M_D{}^{D-2} \int_0^\infty dr \int_0^\infty d^nz 
\int_0^{2\pi}d\theta \,
    \frac{1}{\sigma}\sqrt{\ghat}
\nonumber\\
&\!\!\!=&\!\!\! \frac{2\pi R_0}{n \cdot n! K^{n+1}} M_D{}^{D-2} ~.
\label{WS:reduced-planck-n}
\end{eqnarray}


Let us now consider metric fluctuations around the background of the 
solution~(\ref{WS:metricansatzn}), (\ref{WS:metric-n}), (\ref{WS:solutionnbK}).
As in the last section, we will concentrate only on the linearized
fluctuations of spin 2 modes $h_{\mu \nu}(x^A)$ as in 
Eq.(\ref{WS:metricfluctuation}). 
The linearized Einstein equations are also given by 
Eq.(\ref{WS:linearizedeq2}).
In order to solve the equations, we separate the variables as
\begin{eqnarray}
 h_{\mu\nu}(x^A) = \sum_{m,k,\ell} h^{(mk\ell)}_{\mu\nu}(x^\alpha) 
\phi_{(mk)}(r) \chi_{(k\ell)}(z) e^{i\ell\theta} ~,
\end{eqnarray}
where $\ell$ is an integer angular momentum.
Then, the wave equation is separated into 
\begin{eqnarray}
 -\eta^{\alpha\beta} \partial_\alpha \partial_\beta 
h^{(mk\ell)}_{\mu\nu} (x^\alpha) 
&\!\!\!=&\!\!\! m^2 h^{(mk\ell)}_{\mu\nu}(x^\alpha)
~, \\
 -\partial_r \left( \sigma^2 \partial_r \phi_{(mk)} \right) 
+ k^2 \sigma^2 \phi_{(mk)} (r)
&\!\!\!=&\!\!\! m^2 \sigma \phi_{(mk)} (r)
~, 
\label{WS:phieqn} \\
 -\partial_j \left( \Omega^{n-1} \partial_j \chi_{(k\ell)} (z) \right) 
+ \frac{\ell^2}{R_0{}^2} \Omega^{n-1} \chi_{(k\ell)} (z)  
&\!\!\!=&\!\!\! k^2 \Omega^{n+1} \chi_{(k\ell)} (z)  ~,
\label{WS:chieqn}
\end{eqnarray}
where $\partial_j$ denotes partial differentiation with respect to $z_j$, 
and two eigenvalues $k^2$ and $m^2$ are introduced. 
It is worth emphasizing that the mode equation (\ref{WS:phieqn}) 
for $\phi_{(mk)} (r)$ is identical to the previous case (\ref{WS:phieq2}). 
The differential operator in Eqs.(\ref{WS:chieqn}) is 
self-adjoint if we impose the following boundary conditions\footnote{
We choose the Neumann boundary condition by the same reason as
 the $n=1$ case in the previous section. 
 We should have in mind that this boundary condition at $z_j=\infty$ 
 may be modified when the strong gravity effects are taken into account. 
},
\begin{eqnarray}
 &\!\!\! &\!\!\! \partial_j \chi_{(k \ell)} (z) |_{z_j=0} 
= \partial_j \chi_{(k\ell)} (z) |_{z_j=\infty} = 0 ~.
\label{WS:selfadjointchin} 
\end{eqnarray}
With these conditions we can require the mode 
$\chi_{(k\ell)}(z)$ 
to satisfy the orthonormal condition 
\begin{eqnarray}
 \int_0^\infty d^nz \, \Omega^{n+1} (z)  \chi_{(k \ell)}^* (z)  
\chi_{(k' \ell)} (z)  
&\!\!\!=&\!\!\! \delta_{k k'} ~.
\label{WS:orthonormalchin} 
\end{eqnarray}
Then we obtain  from Eq.(\ref{WS:chieqn}) 
\begin{eqnarray}
 k^2 &\!\!\!=&\!\!\! \int_0^\infty d^nz \, \Omega^{n-1} 
\left| \partial_j \chi_{(k\ell)} \right|^2 
 + \frac{\ell^2}{R_0{}^2} \int_0^\infty d^nz \, \Omega^{n-1} 
\left| \chi_{(k\ell)} \right|^2 \ge 0 
~.
\end{eqnarray}
The equality holds if and only if $\ell =0$ and 
$\chi_{(00)}(z)$ is the zero mode (independent of $z_j$). 
{}For $\phi_{(m k)}(r)$, the boundary conditions (\ref{WS:selfadjointphi}), 
orthonormal condition (\ref{WS:orthonormalphi}), and positivity condition 
(\ref{WS:msquare}) still apply as before. 
Therefore $m^2 \ge 0$ and the equality holds if and only if $k =0$ and 
$\phi_{(00)}(r)$ is a zero mode (independent of $r$).  
These results 
tell us that there is only one massless spin 2 mode $m=k=\ell=0$ 
 of graviton, $h^{(000)}_{\mu\nu}(x^\alpha)$. 
The zero mode wave function in the extra dimensions is 
\begin{eqnarray}
 \chi_{(00)}={\rm const.} ~, \quad \phi_{(00)}={\rm const.} ~,
\end{eqnarray}
The zero mode is localized on the four-dimensional intersection region, 
since the probability distribution in the extra dimensions is strongly 
peaked around the intersection $r=0, z_j=0$ similarly to the case in the 
previous section. 
Thus the zero mode corresponds to the massless graviton localized at 
 the four-dimensional intersection. 
Moreover, we can at least work out all the massive graviton 
fields $h_{\mu \nu}^{(m00)}(x^\alpha)$ with $k=\ell=0$, 
since we know  
the zero mode $\chi_{(00)}(z)$ for $k=\ell=0$ and we have already 
solved the mode equation for $\phi_{(m0)}(r)$ in 
the previous section. 
Therefore we find that there is 
 a continuum 
of massive four-dimensional fields $h^{(m00)}_{\mu\nu}(x^\alpha)$ 
starting 
from $m=0$ similarly to the previous case (and to the original model in 
Ref.\cite{th/9906064}), even though we have not worked out the full 
spectrum of massive modes corresponding to nonvanishing values of $k$. 

The exchange of the massless four-dimensional graviton between two 
sources $\bar m_1, \bar m_2$ on the 
four-dimensional intersection with a large distance $R$ 
gives the usual Newton potential 
(\ref{WS:newtonpotential}) 
also in the present case of $n$ walls. 
Although we have not worked out the massive modes with nonvanishing $k$ 
in detail, we have at least corrections coming from 
the continuum of massive fields $h^{(m00)}_{\mu\nu}(x^\alpha)$ 
starting from $m=0$. 
Since these wave functions are identical to the previous case of the 
seven-dimensional spacetime with $n=1$, we find that there is at least 
the same correction term $\Delta V(R)$ for the Newton potential 
as the previous $n=1$ case (\ref{WS:correctionpotential2}). 
If it turns out that there is a gap in the spectrum of $k$ as 
in the $n=1$ case, all the additional contributions are 
exponentially suppressed. 
Then the above correction term is the leading 
contribution to the correction of the Newton's law of gravitational 
forces. 


\renewcommand{\thesubsection}{Acknowledgments}
\subsection{}

The authors thank Tsuyoshi Sakata for his participation and 
contributions in the early stages of this investigation. 
One of the authors (N.S.) acknowledges a useful discussion with 
Csaba Csaki on the scalar modes and other issues. 
This work is supported in part by Grant-in-Aid 
for Scientific Research from the Japan Ministry 
of Education, Science and Culture for 
the Priority Area 291 and 707.

\end{document}